\documentclass[runningheads]{svmult}

\usepackage{makeidx}   % allows index generation
\usepackage{graphicx}  % standard LaTeX graphics tool
\usepackage{subeqnar}  % subnumbers individual equations
\usepackage{multicol}  % used for the two-column index
\usepackage{physprbb}  % modified textarea for proceedings,
\makeindex             % used for the subject index
\usepackage{enumerate}
\usepackage{psfig}
\usepackage{epsfig}
\newcommand{\scri}{\mycal{I}}

\newcommand{\DIII}{\nabla}
\newcommand{\tDIII}{\tilde{\nabla}}
\newcommand{\DeIII}{\Delta}
\newcommand{\RIII}{\,{}^{\scriptscriptstyle(3)\!\!\!\:}R}
\newcommand{\tRIII}{\,{}^{\scriptscriptstyle(3)\!\!\!\:}\tilde{R}}
\newcommand{\ROI}{\,{}^{\scriptscriptstyle(0,1)\!\!\!\:}\hat R}
\newcommand{\RII}{\,{}^{\scriptscriptstyle(1,1)\!\!\!\:}\hat R}
\newcommand{\epsIII}{\,{}^{\scriptscriptstyle(3)\!\!\!\:}\epsilon}

\newcommand{\hR}{\mbox{$\hat{R}$}}

\DeclareFontFamily{OT1}{rsfs}{}
\DeclareFontShape{OT1}{rsfs}{m}{n}{ <-7> rsfs5 <7-10> rsfs7 <10->
rsfs10}{} \DeclareMathAlphabet{\mycal}{OT1}{rsfs}{m}{n}
\begin{document}
\title*{Problems and Successes in the Numerical Approach 
                      to the Conformal Field Equations}
\toctitle{Problems and Successes in the Numerical Approach
\protect\newline  to the Conformal Field Equations}
% allows explicit linebreak for the table of content
%
%
\titlerunning{Problems and Successes in the Numerical Approach}
% allows abbreviation of title, if the full title is too long
% to fit in the running head
%
\author{Sascha Husa\inst{1}}
\authorrunning{Sascha Husa}
% if there are more than two authors,
% please abbreviate author list for running head
%
%
\institute{Max-Planck-Institut f\"ur Gravitationsphysik, 14476 Golm, Germany}

\maketitle              % typesets the title of the contribution

\begin{abstract}
 This talk reports on the status of an approach to the
 numerical study of isolated systems with the conformal field equations.
 We first describe the algorithms used in a code which has been developed at
 the AEI in the last years, and discuss a milestone result obtained by
 H\"ubner.
 Then we present more recent results as examples to sketch the problems
 we face in the conformal approach to numerical relativity
 and outline a possible roadmap toward making this approach a practical tool.
\end{abstract}

\section{Introduction}\label{sec:intro}
%%%%%%%%%%%%%%%%%%%%%%
Since the early work of Hahn and Lindquist\cite{hahnlindquist64},
Smarr\cite{smarr78} and  Eppley \cite{eppley77}, numerical relativity
has become an important subfield of the theory of gravitation. To outsiders the
progress often seems marginal and unsatisfactory. The classic goal of providing
waveform catalogs for the newly built gravitational wave detectors
has still not been reached (although considerable progress has been made
recently \cite{lazarus}).
By the nature of general relativity, the simulation of isolated systems
poses particularly hard problems. Mathematically such systems can be formalized
by the concept of asymptotically flat spacetimes (see e.g. 
the standard textbook of Wald \cite{Wald}), but it turns out that important
quantities such as the total mass, (angular) momentum or emitted
gravitational radiation can only consistently be defined at
infinity. The traditional approach of introducing an arbitrary spatial cutoff
introduces ambiguities and is not satisfactory at least from a mathematical
point of view. A remedy is suggested by conformal compactification
methods, such as the characteristic approach presented by Luis Lehner
in this volume, or Friedrich's conformal
field equations, which he describes in this volume.
The latter approach avoids the problems associated with the
appearance of caustics in the characteristic formulation
by allowing to foliate the compactified
metric by spacelike hypersurfaces. These hypersurfaces are analogous to the
standard hyperboloid in Minkowski spacetime and
are {\em asymptotically null} in the
physical spacetime. The price to pay is the loss of
the simplicity inherent in the use of null coordinates, and one has to deal
with the full complexity of 3+1 numerical relativity.
%Moreover, Friedrich's judicious reformulation of the equations makes it
%possible to conformally compactify the spacetime in a {\em regular} way.
%Thus, in principle, a global numerical simulation of isolated systems in
%general relativity is possible on finite grids with regular equations.

The fundamental ideas of the numerical solution of the conformal field
equations have been laid out by Frauendiener in this volume, and
in a {\em Living Review} \cite{joerg_livrev},
and he has also discussed his code to treat spacetimes with a
hypersurface-orthogonal Killing vector and toroidal $\scri$'s. 
The purpose of the present article is to show the status of
numerical simulations based on the conformal field equations in 3D --
i.e. three space dimensions without assuming any continuous symmetries --
and to discuss what is needed in order to render this approach %into
a practical tool to investigate physically interesting spacetimes.
By making future null infinity accessible to (completely regular and
well defined) local computations, the approach excels at the extraction of
radiation -- e.g. the quantities to hopefully be measured within the next
years by new large-scale detectors.
One of the main pedagogical goals will be to explain the challenges of
numerical relativity and to highlight some open problems
related to constructing hyperboloidal initial data and actually carrying out
long-time stable and accurate simulations.
For an even more condensed account of the conformal
approach to numerical relativity see \cite{HuAustin}.

The organization is as follows: Sec. \ref{sec:algorithms} introduces
the algorithms developed in the last years by Peter H\"ubner 
(the radiation extraction procedure, which I will only mention briefly,
is based on work of H\"ubner and Marsha Weaver),
and implemented in a set of codes by Peter H\"ubner (who has recently left the
field). All results presented here have been obtained with these codes, which
H\"ubner has described in a series of articles
\cite{PeterI,PeterII,PeterIII,PeterIV}.
Sec. \ref{sec:weakdata} will start with a brief description of the 
evolution of weak initial data which possess a regular point $i^+$
representing future timelike infinity, based on the work of
H\"ubner \cite{PeterIV}.
Then I will discuss the evolution of slightly stronger initial data which
exhibit various problems that will have to be solved, e.g. the choice of gauge,
and use this as a starting point for discussing the main current problems.
In Sec. \ref{sec:computational} purely computational aspects
of this project will be discussed, and in Sec. \ref{sec:discussion} I will
sum up the current status and sketch a possible roadmap for further work.

\section{Algorithms}\label{sec:algorithms}
%%%%%%%%%%%%%%%%%%%%

\subsection{Problem Overview}
%%%%%%%%%%%%%%%%%%%%%

The conformal field equations are formulated in terms of an
unphysical Lorentzian
metric $g_{ab}$ defined on an unphysical manifold ${\cal M}$ which gives
rise to a physical metric $\tilde g_{ab} = \Omega^{-2} g_{ab}$, where the
conformal factor $\Omega$ is determined by the equations. The physical
manifold $\tilde {\cal M}$ is then given by $\tilde {\cal M} =
\{ p \in {\cal M} \, \vert \, \Omega(p) > 0  \}$.

Contrary to the formalism used by Frauendiener in his contribution, we
use a metric based formulation of the conformal field equations:
\label{KonfGl}
\begin{eqnarray}
\label{Riccigleichung}
\nabla_a \hR_{bc} - \nabla_b \hR_{ac} &=& -
 \frac{1}{12} \left( (\nabla_a R) \, g_{bc} - (\nabla_b R) \, g_{ac} \right)
 - 2 \, (\nabla_d \Omega) \, d_{abc}{}^d, \\
\label{Weylgleichung}
\nabla_d d_{abc}{}^d &=& 0, \\
\label{OmGl}
\nabla_a \nabla_b \Omega 
 &=& \frac{1}{4} g_{ab} \, \nabla^c\nabla_c\Omega - \frac{1}{2} \, \hR_{ab} \, \Omega,\\
\label{omWllngl}
\frac{1}{4} \nabla_a \left(\nabla^b\nabla_b \Omega\right) &=& -
\frac{1}{2} \, \hR_{ab} \, \nabla^b \Omega 
- \frac{1}{24} \, \Omega \, \nabla_a R - \frac{1}{12} \, \nabla_a
\Omega \, R ,\\
\label{irrRiemann}
  R_{abc}{}^d &=&  \Omega d_{abc}{}^d
  + ( g_{ca} \hR_b{}^d - g_{cb} \hR_a{}^d 
  - g^d{}_a \hR_{bc} + g^d{}_b \hR_{ac} )/2 \nonumber \\
 & &  + ( g_{ca} g_b{}^d - g_{cb} g_a{}^d ) \frac{R}{12},\\
\label{Rtrace}
  6 \, \Omega \, \nabla^a \nabla_a \Omega &=&
    12 \, (\nabla^a \Omega) \, (\nabla_a \Omega) -   \Omega^2 R.
\end{eqnarray}
Here the Ricci scalar $R$ of $g_{ab}$ is considered a given function
of the coordinates.
For any solution $(g_{ab},\hR_{ab},d_{abc}{}^d,\Omega)$,
$\hR_{ab}$ is the traceless part of the Ricci tensor, and
$\Omega \, d_{abc}{}^d$ the Weyl tensor of~$g_{ab}$.
Note that the equations are regular even for $\Omega=0$.
These ``conformal field equations'' render possible studies of the
global structure of spacetimes, e.g. reading off radiation at null
infinity, by solving regular equations.

The 3+1 decomposition of the conformal geometry can be carried out as usual
in general relativity, e.g.
$$
g_{ab} = h_{ab} - n_{a} n_{b} = \Omega^2 (\tilde h_{ab} - \tilde n_{a}
                                                          \tilde n_{b}),
$$
where $h_{ab}$ and $\tilde h_{ab}$ are the Riemannian 3-metrics
induced by $g_{ab}$ respectively $\tilde g_{ab}$ on a spacelike hypersurface
with unit normals $n_a$, and equivalently $\tilde n_{a} = \Omega n_{a}$
(our signature is $(-,+,+,+)$).
The relation of the extrinsic curvatures ($\tilde k_{ab}=\frac{1}{2}
{\cal L}_{\tilde n}\tilde h_{ab}$, $k_{ab}=\frac{1}{2}
{\cal L}_{n} h_{ab}$) is then easily derived as
$k_{ab} = \Omega (\tilde k_{ab} + \Omega_0 \tilde h_{ab})$, where
$\Omega_0 = n^a \nabla_a \Omega$. Note that for regular components of
$h_{ab}$ and $k_{ab}$, the corresponding components of $\tilde h_{ab}$ and
$\tilde k_{ab}$
with respect to the same coordinate system will in general diverge 
due to the compactification effect. However for the coordinate independent
traces $k= h^{ab} k_{ab}$, $\tilde k = \tilde h^{ab} \tilde k_{ab}$ we get
$$
\Omega k = (\tilde k + 3 \Omega_0),
$$
which can be assumed regular everywhere.
Note that at $\scri$, $\tilde k = - 3 \Omega_0$. Since $\scri^+$ is an ingoing
null surface (with $(\nabla_a \Omega) (\nabla^a \Omega) = 0$ but
$\nabla_a \Omega \neq 0$ ), we have that $\Omega_0 < 0$ at  $\scri^+$.
It follows that $\tilde k > 0$ at $\scri^+$.
We will thus call regular spacelike hypersurfaces in ${\cal M}$ hyperboloidal
hypersurfaces, since in $\tilde {\cal M}$ they are analogous to the standard
hyperboloids $t^2 - x^2 - y^2 - z^2 = \tilde k^2$ in Minkowski space,
which provide the standard example.
Since such hypersurfaces cross $\scri$ but are
everywhere spacelike in ${\cal M}$, they allow to access $\scri$ and
radiation quantities defined there by solving a Cauchy problem (in contrast
to a characteristic initial value problem which utilizes a null surface
slicing). 
Note that hyperboloidal hypersurfaces which cross $\scri^+$ are only Cauchy
surfaces for the {\em future}
domain of dependence of the initial slice of $\tilde {\cal M}$, we therefore
call our studies {\em semiglobal}.

We will not discuss the full $3+1$ equations here for brevity, but rather refer
to \cite{PeterI}. What is important, is that the equations
split into symmetric hyperbolic evolution equations plus constraints
which are propagated by the evolution equations  \cite{PeterI}.
The evolution variables are
$h_{ab}$,
$k_{ab}$,
         the connection coefficients
$\gamma^a{}_{bc}$,
         projections  
$\ROI_a = n^{b}{h_a}^c \hat R_{bc}$ and
$\RII_{ab} = {h_a}^c {h_b}^d \hat R_{bd}$ of the traceless 4-dimensional
             Ricci tensor $\hat R_{bd}$,
         the electric and magnetic components of the
         rescaled Weyl tensor ${d_{abc}}^{d}$
$E_{ab}$, $B_{ab}$, as well as
$\Omega$, $\Omega_0$, $\nabla_a \Omega$, $\nabla^a \nabla_a \Omega$
-- in total this makes $57$ quantities.
In addition the gauge source functions $q$, $R$ and $N^a$ have to be specified,
in order to guarantee symmetric hyperbolicity they are given as functions of
the coordinates.
Here $q$ determines the lapse 
as $N= e^q \sqrt {\det h}$ and $N^a$ is the shift vector. The Ricci scalar $R$
can be thought of as implicitly steering the conformal factor $\Omega$.

The numerical treatment of the constraints and evolution equations will be
described below. But before, let us spend some time
on general considerations about the treatment of null infinity.
Since $\Omega$ is an evolution variable and not specified a priori,
$\scri$ will in general not be aligned with grid points, and
interpolation has to be used to evaluate computed quantities at locations
of vanishing $\Omega$. For the physically interesting case
of modeling an isolated system, ``physical''
$\scri$ -- i.e. the component of $\scri$ that idealizes us outside observers
and our gravitational wave detectors (neglecting cosmological effects such as
redshift etc.) -- has spherical topology.
There may be more than one component of $\scri$, i.e. additional spherical
components associated with ``topological black holes'' (see Sec.
\ref{subsec:bh}).
In principle it is possible of course to control the movement of $\scri$
through the grid by the gauge choice --
see \cite{ScriFixing} for how to achieve
such $\scri$ {\em fixing} within Frauendiener's formulation.
An example would be the so called
$\scri$ {\em freezing}, where $\scri$ does not change its coordinate
location.

What is the significance of how $\scri$ moves through the grid? 
This question is directly related to the question for the global structure of
spacetime. Although  many questions are left open, the present understanding
of the global structure of generic vacuum spacetimes, which can be constructed
from regular initial data, does provide some hints.
First, note that spacetimes which are asymptotically flat in spacelike and null
directions -- i.e. isolated systems -- do not necessarily have to be
asymptotically flat in timelike directions. An example would be a spacetime
that contains a star or a black hole. In such cases where the end state of the
system is not flat space, we can not expect the conformal spacetime
$({\cal M}, g_{ab})$ to contain a regular point $i^+$.
In the case of sufficiently weak data however, Friedrich has shown in
\cite{Fr86ot}
that a regular point $i^+$ will exist\label{fig:slicings}
 -- consistent with our intuition.
The global structure is then similar to Minkowski space.
The standard conformal compactification of Minkowski space
is discussed in textbooks (see e.g. \cite{Wald}) as a mapping to the
Einstein static universe.
There $\scri$ moves inward and contracts to a point within finite coordinate
time. In order to resolve such situations it seems most appropriate to
choose a gauge which mimics this behavior, i.e. where $\scri$ contracts to
a point after finite coordinate time.
The boundary of the computational domain is set in the unphysical
region and the physical region contracts in coordinate space.
Accordingly, the initial data are also extended beyond the physical region of 
spacetime. 
It is this scenario which is best understood so far, and which is presented
in some more detail in Sec. \ref{sec:weakdata}. 

For sufficiently strong regular data it is known
that singularities develop \cite{Pe63ap} -- according to the cosmic
censorship conjecture \cite{cosmic-censorship}
such singularities should generically
be hidden inside of black holes. For such data
we cannot expect a regular point $i^+$ to exist. In the case when
$i^+$ is singular
(and not much else is currently known even about the $i^+$ of Kruskal spacetime -- see however Bernd Schmidt's contribution in this volume) we 
have to expect structure like sharp gradients near $i^+$, which makes it
unlikely that we can afford to significantly reduce the size of the physical
region in coordinate space (at least not without adaptive mesh refinement --
a technology not yet available for 3D evolutions). A $\scri$-freezing gauge
may be appropriate for such a situation. Furthermore, phenomena like
quasi-normal ringdown, or the orbital motion of a two-black
hole system suggest that the numerical time-coordinate 
better be adapted to the intrinsic time scale of the system. Associated
with quasi-normal ringdown for example is a fixed period in Bondi-time, which
suggests Bondi time as a time coordinate near $\scri$ in a situation
dominated by ringdown.
Thus, for  black hole spacetimes it might turn out that the best choice of
gauge fixes $\scri$ to a particular coordinate position, and shifts
$i^+$ into the infinite future \footnote{Note however, that
in the absence of results, we are left to speculation here.}.
It could be possible that in such a case the boundary can be chosen to
either coincide with $\scri$, or to be put just a small number of gridpoints
outside, which
would raise the question for an evolution algorithm that does not require a
topologically rectangular grid.
We see that the optimal choice of numerical algorithms and gauges may be
tightly related to the global structure of the investigated spacetimes --
which is actually one of the main questions our simulations should be able to
answer!

\subsection{Construction of ``extended'' hyperboloidal initial data}
%%%%%%%%%%%%%%%%%%%%%%%%%%%%%%%%%%%%%%%%%%%%%%%%%%%%%%%%%%%%%%%%

Evolution of a solution to the Einstein equations 
starts with a solution to the constraints.
The constraints of the conformal field equations (see Eq. (14) of Ref.
\cite{PeterI}) are regular equations on the whole conformal spacetime
$({\cal M}, g_{ab})$.
However, they have not yet been cast into a standard type of PDE system,
such as a system of elliptic PDEs \footnote{work toward this goal
is reported in this volume by Adrian Butscher.}.
One therefore resorts to a 3-step method \cite{PeterIII}:
\begin{enumerate}
\item Obtain data for the Einstein equations: the first and
second fundamental forms $\tilde h_{ab}$ and ${\tilde k}_{ab}$ induced
on $\tilde\Sigma$ by $\tilde g_{ab}$, corresponding in
the compactified picture to $h_{ab}$, ${k}_{ab}$ and 
$\Omega$ and $\Omega_0$.
 This yields so-called ``minimal data''.
\item  Complete the minimal data on $\bar\Sigma$  to data for {\em all}
 variables using the conformal
constraints -- {\em in principle} this is mere algebra and
differentiation.
\item Extend the data from $\bar\Sigma$ to $\Sigma$ in some ad hoc but
sufficiently smooth and ``well-behaved'' way.
\end{enumerate}

In order to simplify the first step, the implementation of the code is 
restricted to a subclass of
hyperboloidal slices where initially ${\tilde k}_{ab}$ is pure trace,
${\tilde k}_{ab} = \frac{1}{3} {\tilde h}_{ab} \tilde k$.
The momentum constraint
$
  \tDIII^b {\tilde k}_{ab} - \tDIII_a \tilde k = 0  
$
then implies $\tilde k = \mbox{const.} \ne 0$. We always set $\tilde k > 0$.
In order to reduce the Hamiltonian constraint
$$
\label{HamilConstr}
  \tRIII + {\tilde k}^2 = {\tilde k}_{ab}{\tilde k}^{ab}
$$
to {\em one} elliptic equation of second order,
we use a modified Lichnerowicz ansatz 
$$
  \label{tildeh}
  {\tilde h}_{ab} = \bar\Omega^{-2} \phi^4 h_{ab}
$$
with {\em two} conformal factors $\bar\Omega$ and $\phi$. The principal idea
is to choose  $h_{ab}$ and $\bar\Omega$, and solve for $\phi$, as we will
describe now.
First, the ``boundary defining'' function $\bar\Omega$ is chosen to
vanish on a 2-surface ${\cal S}$ -- the boundary of $\bar\Sigma$
and initial cut of $\scri$ -- with non-vanishing gradient on ${\cal S}$.
The topology of  ${\cal S}$ is chosen as spherical for asymptotically
Minkowski spacetimes. 
Then we choose $h_{ab}$ to be a Riemannian metric on $\Sigma$,
with the only restriction
that the  extrinsic 2-curvature induced by $h_{ab}$ on ${\cal S}$ is pure
trace, which is required as a smoothness condition \cite{AnCA92ot}. 
With this ansatz ${\tilde h}_{ab}$ is singular at ${\cal S}$,
indicating that ${\cal S}$ represents an infinity.
The Hamiltonian constraint then reduces to the Yamabe equation for the
conformal factor $\phi$:
$$
  4 \, \bar\Omega^2 \DIII^a \DIII_a\phi
  - 4 \, \bar\Omega (\!\DIII^a \bar\Omega)(\!\DIII_a \phi)
  - \left( \frac{1}{2} \RIII \, \bar\Omega^2 + 2 \bar\Omega \DeIII\bar\Omega
           - 3 (\!\DIII^a \bar\Omega) (\!\DIII_a \bar\Omega) 
    \right) \phi
  = \frac{1}{3} {\tilde k}^2 \phi^5.
$$
This is a semilinear elliptic equation -- except at ${\cal S}$,
where the principal part vanishes for a regular solution. This
however determines the boundary values as
\begin{equation}\label{boundaryvals}
\phi^4 =9 \, {\tilde k}^{-2} \, (\nabla^a \bar{\Omega}) \,
 (\nabla_a \bar{\Omega}).
\end{equation}
Existence and uniqueness of a positive solution to the Yamabe 
equation and the corresponding existence and uniqueness of regular data for
the conformal field equations using the approach outlined above have been
proven by Andersson, Chru\'sciel and Friedrich~\cite{AnCA92ot}.
Solutions to the Yamabe equation -- and thus minimal initial data --
can either be taken from exact solutions or from numerical solutions of the
Yamabe equation. Exact solutions which possess a $\scri$ of spherical
topology have been implemented for Minkowski space and for
Kruskal spacetime -- see the contribution of Bernd Schmidt in this volume.
These solutions are defined even outside of $\scri$,
and thus can directly be completed to initial data
for all variables by using the conformal constraints.

If the Yamabe equation is solved numerically, the boundary has to be chosen
at  ${\cal S}$, the initial cut of $\scri$,
with boundary values satisfying Eq. (\ref{boundaryvals}).
If the equation would be solved on a larger (more convenient
Cartesian) grid, generic boundary conditions would cause the solution to
lack sufficient differentiability at ${\cal S}$,
see H\"ubner's discussion in \cite{PeterII}. This problem is due to
the degeneracy
of the Yamabe equation at $\cal S$. Unfortunately, this means
that we have to solve an elliptic problem with {\em spherical boundary}.
This problem is solved by combining the use of spherical coordinates with
pseudo-spectral collocation methods. In pseudo-spectral methods the solution
is expanded in
(analytically known) basis functions -- here a Fourier series for the angles
and a Chebychev series for the radial coordinate. 
For an introduction to  pseudo-spectral methods see e.g.
\cite{Fornberg,Canuto,Boyd}.
This allows to take care of coordinate singularities in a clean way, provided
that all tensor components are computed with respect to a regular
(e.g. Cartesian) basis and that no collocation points align with the
coordinate singularities.  Another significant advantage of spectral methods
is their fast convergence: for smooth solutions they typically converge
exponentially with resolution. 
The necessary conversions between the collocation and spectral representations
are carried out as fast Fourier transformations with the FFTW library
\cite{FFTW}.
The nonlinearities are dealt with by a
Newton iteration \cite{NewtonIteration}, the resulting linear equations are
solved
by an algebraic multigrid linear solver (the AMG library \cite{AMG}).

The constraints needed to complete minimal initial data to data for 
all evolution variables split into two groups: those that require divisions
by the conformal factor $\Omega$ to solve for the unknown variable,
and those which do not.
The latter do not cause any problems and can be solved without
taking special care at $\Omega = 0$.
The first group, needed to compute $\RII$, $E_{ab}$ and $B_{ab}$,
however does require special numerical
techniques to carry out the division,
and furthermore it is not known if their solution outside of
$\scri$ actually allows solutions which are sufficiently smooth
beyond $\scri$. Thus, at least for these we
have to find some ad-hoc extension. Note that 
in the case of analytical minimal data, the additional constraints are
solved on the whole time evolution grid.

The simplest approach to the division by $\Omega$ would be an implementation of
l'Hospital's rule, however this leads to nonsmooth errors and consequently
to a loss of convergence \cite{PeterII}.
Instead H\"ubner \cite{PeterII} has developed a technique to replace
a division
$g = f/\Omega$ by solving an elliptic equation of the type (actually
some additional terms added for technical reasons are omitted here for
simplicity)
$$
   \DIII^a \DIII_a ( \Omega^2 g - \Omega f ) = 0 
$$
for $g$.
For the boundary values $\Omega^2 g - \Omega f = 0$, the unique solution is
$g=f/\Omega$. For technical details see \cite{PeterIII}.
The resulting linear elliptic equations for $g$ are solved by the same 
numerical techniques as  the Yamabe equation.
For technical details see H\"ubner \cite{PeterIII}.

Finally, we have to extend the initial data to the full Cartesian
spatial grid in some way.
Since solving all constraints also outside of $\scri$ will in general not be
possible in a sufficiently smooth way, we have to find an ad hoc extension,
which violates the constraints outside of $\scri$ but is sufficiently well
behaved to serve as initial data. The resulting constraint violation is not
necessarily harmful for the evolution, since $\scri$ causally disconnects
the physical region from the region of constraint violation. On the numerical
level, errors from the constraint violating region {\em will} in general
propagate into the physical region, but if our scheme is
consistent, these errors have to converge
to zero with the convergence order of the numerical scheme
(fourth order in our case). There may still be practical problems, that
prevent us from reaching this aim, of course: making the ad-hoc extension
well behaved is actually quite difficult, the initial constraint violation may
trigger constraint violating modes in the equations, which take us away from 
the true solution, singularities may form in the unphysical region, etc.
%Most of the calculations have used a $C^2$ extension for all variables, which
%is based on a  $C^2$ coordinate transformation for the radial Chebychev series.
%Recently $C^{\infty}$ extensions have been implemented at the cost of more
%parameters to be tuned.

Since the time evolution grid is Cartesian, its grid points will in general not
coincide with the collocation points of the pseudo-spectral grid. Thus
fast Fourier transformations can not be used for transformation to
the time evolution grid. The current implementation instead uses
standard discrete (``slow'') Fourier transformations, which typically take up
the major part of the
computational effort of producing initial data. 

It turns out, that the combined procedure works reasonably well for 
certain data sets. For other data sets the division by $\Omega$
is not yet solved in a satisfactory way, and constraint violations
are of order unity for the highest available resolutions.
In particular this concerns the constraint
$\DIII_b E_a{}^b = - \epsIII_{abc} k^{bd} B_d{}^c$ (Eq. (14d) in
 \cite{PeterI}),
since $E_{ab}$ is computed last in the hierarchy of variables and
requires two divisions by $\Omega$. Further research
is required to analyze the problems and either improve the current 
implementation or apply alternative algorithms.
Ultimately, it seems desirable to change the algorithm of obtaining
initial data to a method that solves the conformal constraints directly and
therefore does not suffer from the current problems.
This approach may of course introduce new problems like an elliptic system
too large to be handled in practice.

\subsection{Black hole initial data}\label{subsec:bh}
%%%%%%%%%%%%%%%%%%%%%%%%%%%%%%%%%%%%

Since the standard definition of a black hole as the interior of
an event horizon is a global concept, it is a priori not clear what
one should consider as ``black hole initial data''. In practice,
the singularity theorems \cite{Pe63ap} and the assumption of cosmic censorship
\cite{cosmic-censorship} 
usually lead to the identification of
``black hole initial data'' with data that contain apparent horizons,
and to associate the number of apparent horizons with the number of black
holes in the initial data \footnote{Note that both the number of components of
the event and apparent horizon are slice-dependent.}.

A common strategy to produce apparent horizons is to use
topologically nontrivial data, that is data which
possess more than one asymptotically flat region. In the time-symmetric
case such data obviously possess a minimal surface!
Asymptotic ends that extend to spatial infinity $i^0$ are relatively
easy to produce by  compactification methods, see e.g.
\cite{BeOM,Be94Ini,HuDiss,bernd_punctures}
or the contribution of Dain in this volume.
From the numerical point of view it is important that
the topology of the computational grid is independent of the number of
asymptotic
regions or apparent horizons considered: suitable regularization procedures
allow to treat spatial infinities as grid points.

In the current approach to the hyperboloidal initial value problem, where
first the Yamabe equation needs to be solved, the grid topology
{\em does depend} on the number of topological black holes -- in this case
the number of initial cuts of $\scri$'s, which have spherical topology.
One option would be of course to combine both ingredients and consider
``mixed asymptotics'' initial data, which extend to the physical $\scri$
and to unphysical interior spacelike infinities which only serve the purpose
of acting as ``topological sources'' for apparent horizons.

Another option, suggested by H\"ubner in \cite{PeterIII},
is to generalize the current
code for the initial data, which only allows for one cut of $\scri$,
which has spherical topology,  to multiple $\scri$'s of spherical topology.
For the case of one black hole this would correspond to the relatively simple
modification to   $S^2 \times R$ topology. For the case of
two black holes one could implement the
Schwarz alternating procedure (as described in Sec. 6.4.1 of
Ref. \cite{QuV97NA}) to treat
three $\scri$'s with three coordinate patches, where each patch is adapted to a
spherical coordinate system with its $\scri$.
A more practical approach (at least to get started) could be to produce
topologically
trivial black hole initial data. Since we expect physical black holes to
result from the collapse of topologically trivial regular initial data,
such data would in some sense be more physical. Theorems on the existence
of apparent horizons in Cauchy data have been presented by
Beig and \'O Murchadha in \cite{BeOM}.
Numerical studies in this spirit have been
performed by the author \cite{HuDiss}.
Such data could in principle be produced
with the current code once it gets coupled to an apparent horizon finder.
For the hyperboloidal initial value problem it is actually not known, whether
such data actually exist, but it seems physically reasonable. Finding such
data numerically by parameter studies would be an interesting result in itself.
A natural question in this context
is whether there is any  qualitative difference
between  ``topological'' and ``non-topological'' black holes outside of
the event horizon, e.g. regarding their waveforms?

\subsection{Numerical setup for evolutions}
%%%%%%%%%%%%%%%%%%%%%%%%%%%%%%%%%%%%%%%%%%%

The time evolution algorithm is an implementation of a standard
fourth order method of lines (see e.g. \cite{GuKA95TD}),
with centered spatial differences and
Runge-Kutta time integration. Additionally, a dissipation term of the
type discussed in theorems ~6.7.1 and ~6.7.2 of  Gustafsson, Kreiss and
Oliger \cite{GuKA95TD} is added to
the right-hand-sides to damp out high frequency oscillations and
keep the code numerically stable.  
%The dissipation term used is  $\sigma Q_2 := \frac{\sigma }{64 \, N} (\Delta x)^5 \sum_{i=1}^{N}
%\partial_i{}^6 f$, where the spatial derivatives are discretized as
%\begin{eqnarray}
%  \lefteqn{
%    \partial_x{}^6 {f}^l_{i,j,k} = 
%      \frac{1}{(\Delta x)^6}
%      \left( {f}^l_{i-3,j,k} - 6 {f}^l_{i-2,j,k}
%             + 15 {f}^l_{i-1,j,k} \right. }
% \nonumber\\
%  & & \left. {}
%             - 20 {f}^l_{i,j,k}
%             + 15 {f}^l_{i+1,j,k} 
%             - 6 {f}^l_{i+2,j,k} + {f}^l_{i+3,j,k}
%    \right). \nonumber        
%\end{eqnarray}
%This dissipation term has been discussed in detail by.
%The dissipation term contains an amplitude parameter $\sigma$
%enough and \dots {\bf FIXME: more detail} the algorithm is stable. Note that of course sufficient
%resolution may not be obtainable on a given machine. Also, choosing $\sigma$
%too large will cause spurious results. This makes the adjustment of
%the dissipation term still somewhat of an art, however our 
Numerical experiments show that usually small amounts if dissipation are
sufficient (the dissipation term used contains a free parameter),
and do not change the results in any significant manner.

A particularly subtle part of the evolution usually is the boundary treatment. 
In the conformal approach we are in the situation that the boundary is actually
situated outside of the physical region of the grid -- this is one of
its essential advantages! In typical explicit time evolution algorithms,
such as our Runge-Kutta method of lines, the numerical propagation speed
is actually larger than the speed of all the characteristics (in our case
the speed of light). Thus $\scri$ does {\em not} shield the physical region
from the influence of the boundary  -- but this influence has to converge
to zero with the convergence order of the algorithm --
fourth order in our case. One therefore does not have to choose a ``physical''
boundary condition, the only requirements are stability and ``practicality'' --
e.g. the boundary condition should avoid, if possible, the development of
large gradients in the unphysical region to reduce the numerical 
``spill over'' into the physical region, or even code crashes.

The current implementation relies on a ``transition layer''
in the unphysical region, which is used to transform  the rescaled
 Einstein equations to trivial evolution equations, which are stable with
a trivial copy operation at outermost gridpoint as a boundary condition
(see Ref. \cite{PeterI} for details and references).
We thus modify the evolution equations according to
replace
$$
  {  \partial_t {f} 
  + {{A}}^i \partial_i {f} - 
  {b} = 0}
\quad
\rightarrow
\quad
  { \partial_t {f} + \alpha(\Omega) \left( {{A}}^i \partial_i {f} -
  {b} \right) = 0,}
$$
where $\alpha$ is chosen as $\alpha(\Omega) = 0$ for
$\Omega\leq\Omega_0<\Omega_1<0$ and $1$ for
$\Omega\geq\Omega_1$. 
This procedure works reasonably well for weak data, however there are some
open problems. One is, that the region of large constraint violations outside
of $\scri$ may trigger constraint violating modes of the equations that can
grow exponentially. Another problem ist that a ``thin'' transition zone causes
large gradients in the coefficients of the equations -- thus eventually leading
to large gradients in the solution, while a ``thick'' transition zone
means to loose many gridpoints. If no transition zone is used at all, and the
Cartesian grid boundary touches $\scri$, the ratio of the number of 
grid points in the physical region versus the number of 
grid points in the physical region is already $\pi/6 \approx 0.52$.

\subsection{Physics Extraction}
%%%%%%%%%%%%%%%%%%%%%%%%%%%%%%%%

Extracting physics from a numerical solution to the Einstein equations
is a nontrivial task. Results typically show a combination of physics
and coordinate effects which are hard to disentangle, in particular in the
absence of a background geometry or preferred coordinate system.
In order to understand what is going on in a simulation, e.g. to find ``hot
spots'' of inaccuracy or instability or bugs in an algorithm,
it is often very important to visualize the ``raw'' data of a calculation.
Here the visualization of scalar and in particular tensor fields in
3D is a subtle task in itself. But beyond that, one also
wants ways to factor out coordinate effects in some way, and ideally 
access physical information directly.

One way commonly used to partially factor our coordinate effects is to look at
curvature invariants, another possibility is to trace geodesics through
spacetime.
In the current code this is done by concurrent integration of geodesics by
means of the same 4th order Runge-Kutta scheme used already in the method of
lines.
Both null and timelike geodesics, as well as geodesics of the physical
and rescaled metrics can be computed, and various quantities such as curvature
invariants are computed by interpolation along the geodesics.

\begin{figure}[htbp] % [b!] fig 1
\centerline{\epsfig{file=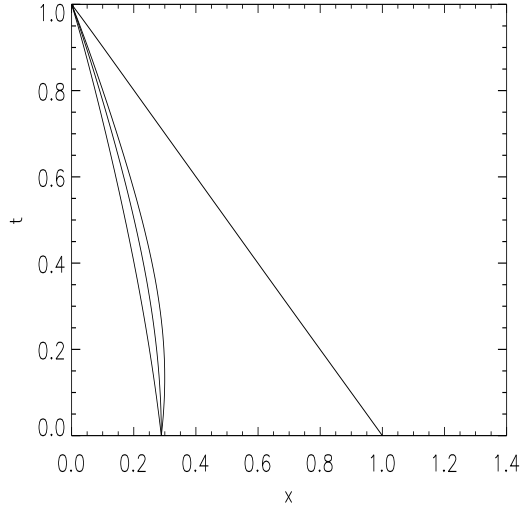,height=3.01in,width=3.01in}}
\vspace{10pt}
\caption{Three timelike geodesics of different initial velocities
 (starting at $x=\frac{1}{2\sqrt{3}}$) meet a generator of $\scri$
 (starting at $x=1$) at future timelike infinity $i^+$.}
\label{MeetingPoint}
\end{figure}

Particularly important are null geodesics propagating along $\scri$, since they
can be used to define a Bondi system and thus compute radiation quantities
such as the Bondi mass or news.
Note that the foliation of spacetime chosen for evolution will in general
{\em not} reproduce cuts of $\scri$ of constant Bondi time.
H\"ubner has therefore implemented postprocessing algorithms
(using the IDL programming language/software system)
which construct slices of constant Bondi time in the data corresponding to
the null geodesics propagating on $\scri$ by interpolation
(the algorithms are based on unpublished work of H\"ubner and Weaver).
This evolution of geodesics is illustrated by Fig. \ref{MeetingPoint},
which shows three timelike geodesics
originating with different initial velocities at the same point
$(x_0,y_0,z_0)=(\frac{1}{2\sqrt{3}},\frac{1}{2\sqrt{3}},\frac{1}{2\sqrt{3}})$
meeting a generator of $\scri$ at $i^+$. 

%An important tool to analyze numerically constructed spacetimes
%is an apparent horizon finder, e.g. to 

\section{Results for weak data}\label{sec:weakdata}
%%%%%%%%%%%%%%%%%%%%%%%%%%%%%%%

In this section I will discuss results of 3D calculations for initial data 
which evolve into a regular point $i^+$, and which thus could be called 
``weak data''. Bernd Schmidt presents results for the Kruskal spacetime
in this volume (see also \cite{bernd_kruskal}).
The initial conformal metric is chosen in Cartesian coordinates as
\begin{equation}\label{eq:standard-h}
ds^2 =  \left(1 + \frac{A}{3}
   \bar\Omega^2 \left( x^2 + 2 y^2 \right)\right) dx^2 + dy^2 + dz^2.
\end{equation}
The boundary defining function is chosen as
$\bar\Omega = \left( 1 - \left(x^2+y^2+z^2\right) \right)/2,$
it is used to satisfy the smoothness
condition for the conformal metric at $\scri$.

These data have been evolved previously by H\"ubner for $A=1$ as reported
in \cite{PeterIV}. For the gauge source functions
he has made the ``trivial'' choice:
$R=0$, $N^a = 0$, $q=0$, i.e. the conformal spacetime has vanishing scalar
curvature, the shift vanishes and the lapse is given by
$ N = e^{q} \sqrt{\det h} =\sqrt{\det h}$.
This simplest choice of gauge is completely sufficient for $A=1$ data, and has
lead to a milestone result of the conformal approach --
the evolution of weak data which evolve into a  regular point $i^+$
of $\cal M$, which is resolved as a single grid cell.
With this result H\"ubner has illustrated a theorem by Friedrich,
who has shown that for sufficiently weak initial data there exists a 
regular point $i^+$ of $\cal M$ \cite{Fr86ot}.
The complete future of (the physical part of) the initial slice
can thus be reconstructed in a finite number of computational time steps.
This calculation is an example of a situation for
which the usage of the conformal field equations is ideally suited:
main difficulties of the problem are directly addressed and solved
by using the conformal field equations.

\begin{figure}[htbp] % [b!] fig 1
\centerline{\epsfig{file=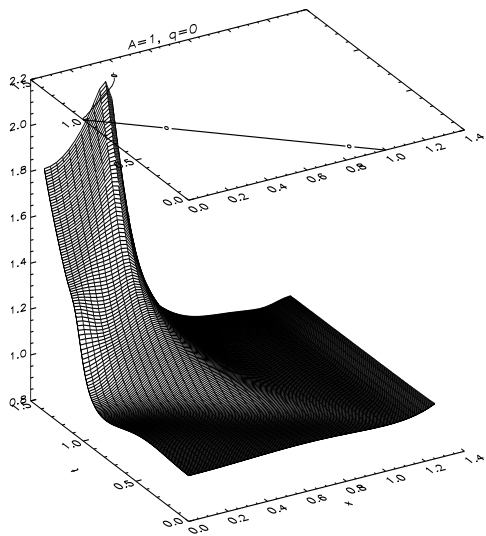,height=3.01in,width=3.01in}}
\vspace{10pt}
\caption{
Lapse $N$ for the $x$-axis, for amplitude $A=1$ and $R=N^a=q=0$.
The contour line marks $\scri$
($\Omega=0$).}
\label{Lapse-N1}
\end{figure}

\begin{figure}[htbp] % [b!] fig 1
\centerline{\epsfig{file=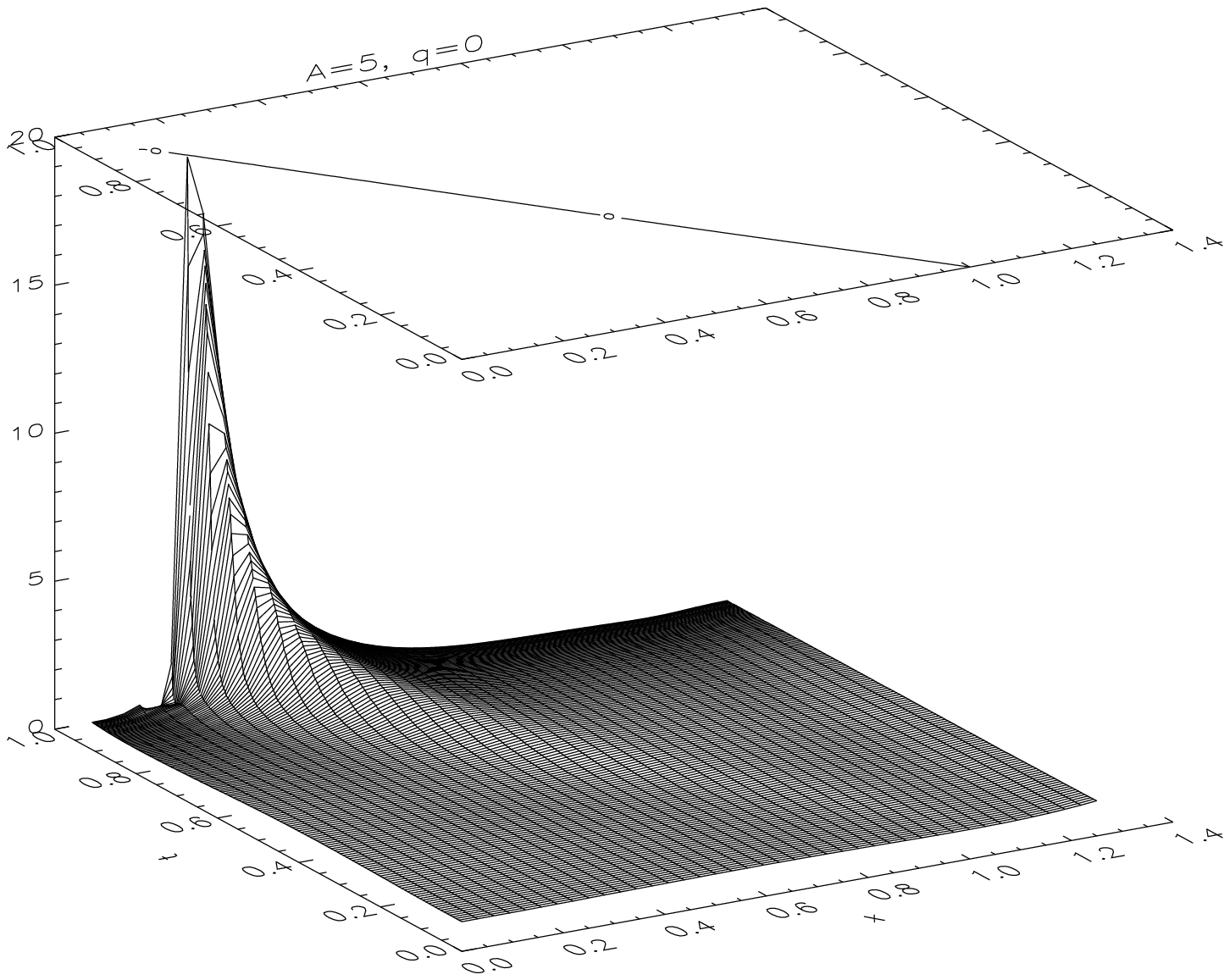,height=3.01in,width=3.01in}}
\vspace{10pt}
\caption{
Lapse $N$ for the $x$-axis, for $A=5$ and $R=N^a=q=0$. The contour line marks $\scri$
($\Omega=0$).}
\label{Lapse-N5}
\end{figure}

The natural next question to ask is: what happens if one increases the
amplitude $A$? To answer this question, I have performed and analyzed
runs for integer values of $A$ up to $A=20$. The results
presented here have been produced with low resolutions of $80^3$
(but for higher or slightly lower resolutions we essentially get the
same results). For convergence tests of the code see 
\cite{PeterII,PeterIII,PeterIV}.
While for $A=1, 2$ the 
code continues beyond $i^+$ without problems, for all higher
amplitudes the ``trivial'' gauge leads to code crashes before reaching $i^+$.
Here by ``code crash'' we mean that computational values get undefined,
e.g. the code produces ``not a number'' (NaN) values.
While the physical data still decay quickly in time, a sharp peak of the lapse
develops outside of $\scri$ and crashes the code after 
Bondi time $\sim 8 (320 M)$ for $A=3$ and $\sim 1.5 (3 M)$ for $A=20$
(here $M$ is the initial Bondi mass).
In Figs. \ref{Lapse-N1} and \ref{Lapse-N5} the lapse $N$
is plotted for runs with $A=1$ and
$A=5$. While for $A=1$ the lapse only shows significant growth after $t=1$
($i^+$ is located at $x=y=z=0$, $t=1$), for $A=5$ a very sharp peak grows
outside of $\scri$ and crashes the code at $t=0.9076$.
Where does this rapid growth come from? Note that the initial
conformal metric Eq. (\ref{eq:standard-h}) shows significant growth outside
of $\scri$. Combined with the lapse $N=\sqrt{\det h}$ this leads to a
growth of the lapse toward the  grid boundary. In the present case a positive
feedback effect with a growth of metric components in time seems to be
responsible for the eventual crash of the code.
Note that this feedback only takes place in a small region outside of
$\scri$ -- further outward it is prevented
by the transition to trivial evolution equations.

\begin{figure}[htbp] % [b!] fig 1
\centerline{\epsfig{file=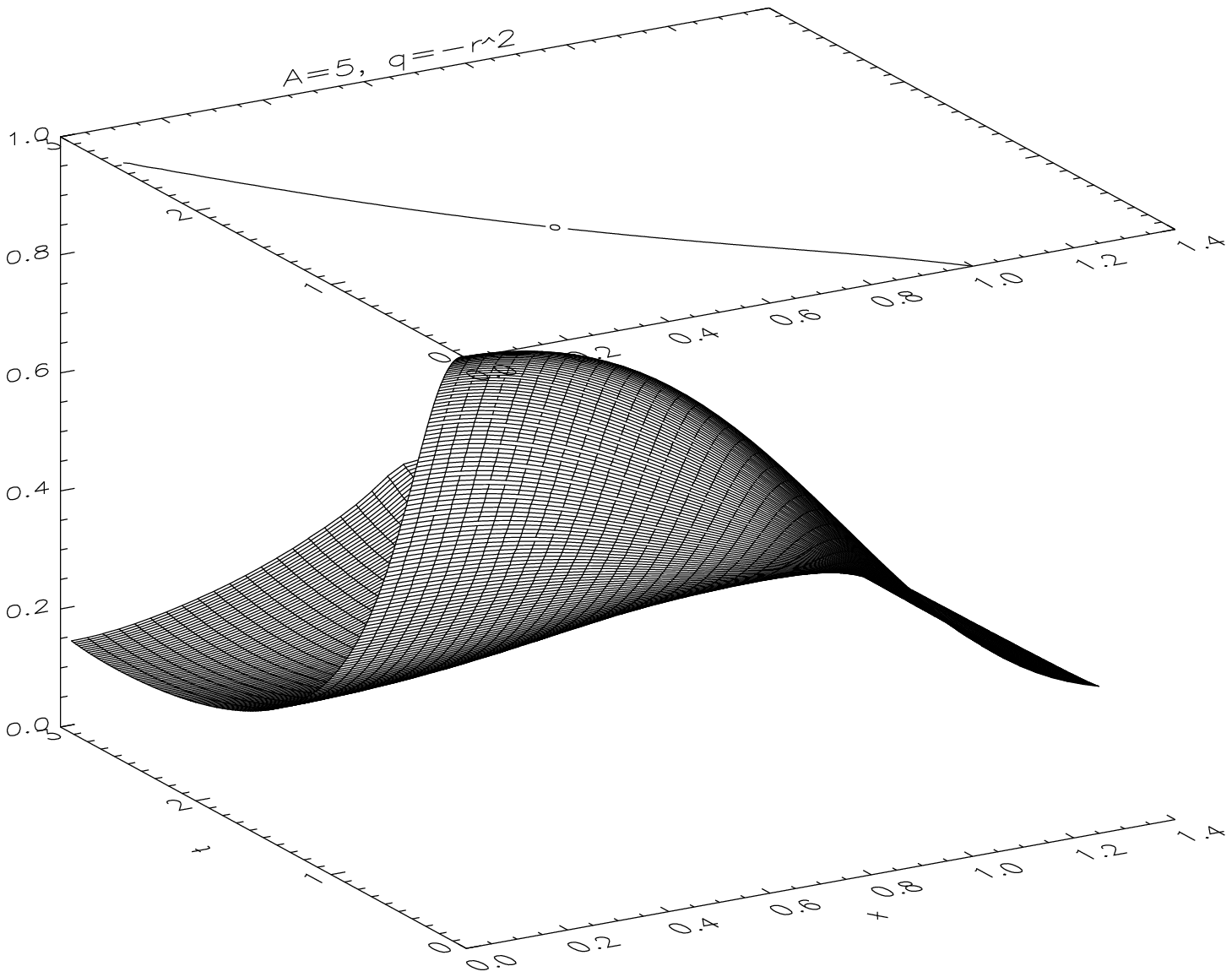,height=3.01in,width=3.01in}}
\vspace{10pt}
\caption{
Lapse $N$ for the $x$-axis, for $A=5$ and $R=N^a=0$, $a=-r^2$. The contour line marks $\scri$
($\Omega=0$). }
\label{Lapse-N5q}
\end{figure}

Fig. \ref{Lapse-N5q} shows a cure of the problem: a modified gauge source
function $q = -r^2/a$ ($N=e^{-r^2/a}\sqrt{\det h}$) with $a=1$ leads to a very
smooth
lapse (and correspondingly also to smooth metric components). Note that in
Fig. \ref{Lapse-N5q}, due to the different lapse,
the point $i^+$ is  {\em not} located at $t=1$. The value of $a=1$
here is found by moderate tuning of $a$ to a best value (significantly
decreasing or increasing $a$ crashes the code before $i^+$ is reached).
Unfortunately, this modification of the lapse is not sufficient to achieve
much higher amplitudes. As $A$ is increased, the parameter $a$ requires
more fine tuning, which was only achieved for $A \leq 8$. For
higher amplitudes the code crashes with significant
differences in the maximal and minimal Bondi time achieved, while
the radiation still decays very rapidly and the news scales
almost linearly. Furthermore, the curvature
quantities do not show excessive growth -- it is thus natural to assume that
we are still in the weak-field regime, and the crash is not connected to
the formation of an apparent horizon or singularity.

These results suggest that in order to model a gauge source function
$q$ that would allow to
evolve up to $i^+$, one would need more that one parameter, e.g. at least
3 parameters for a non-isotropic ansatz such as
$q = - (x^2/a^2 + y^2/b^2 + z^2/c^2)$ or something similar.
To tune 3 or more parameters
for each evolution seems however computationally prohibitive. While some
improvement is obviously possible through simple non-trivial models for
the lapse (or other gauge source functions), this approach seems very limited
and more understanding will be necessary to find practicable gauges.
An interesting line of research would be to follow the lines
of Ref. \cite{Miguel} in order to find evolution equations for the gauge source
functions which avoid the development of pathologies.
A particular aim would be to find equations such that the resulting system
of evolution equations is symmetric hyperbolic.

\begin{figure}[htbp] % [b!] fig 1
\centerline{\epsfig{file=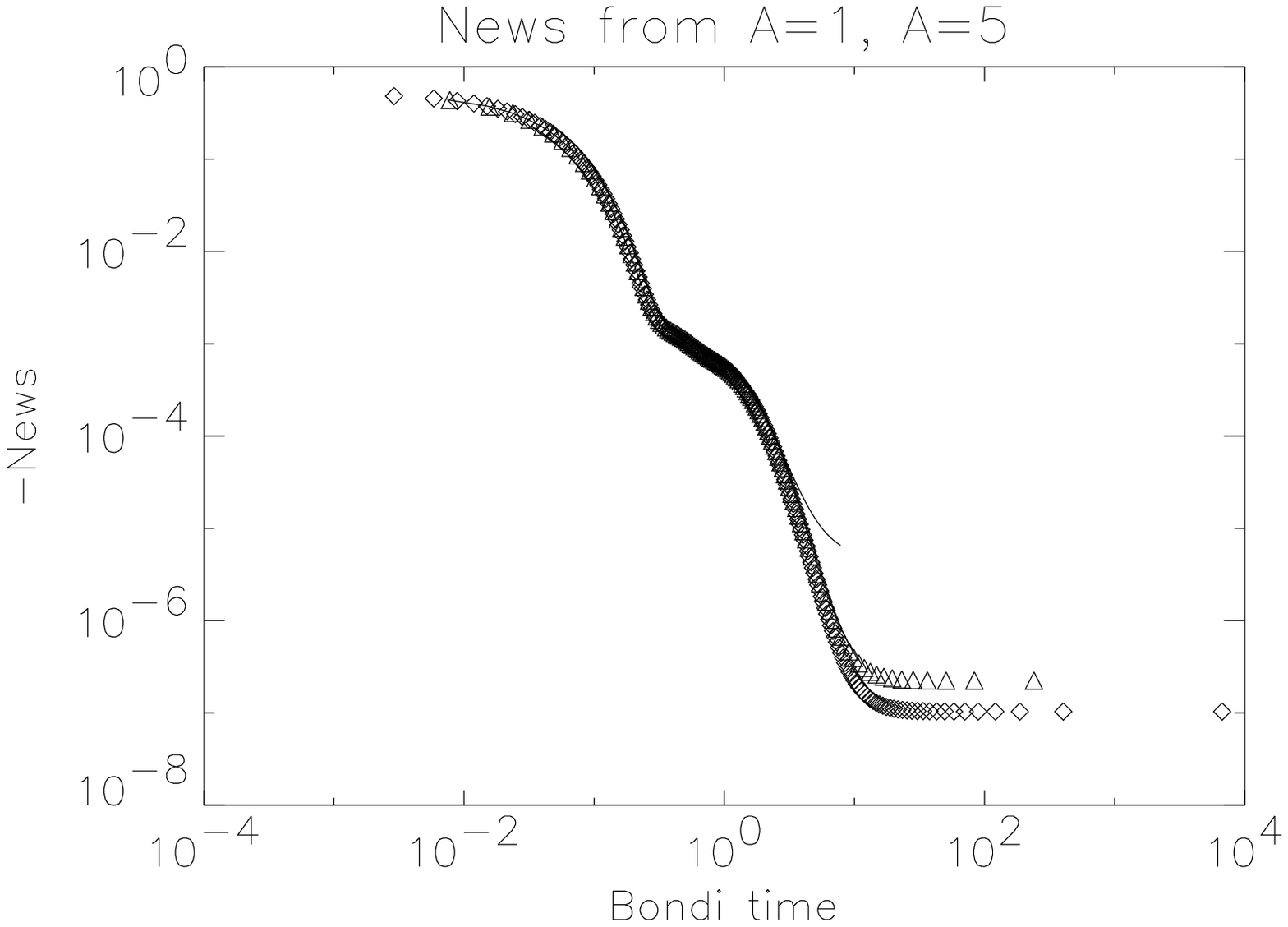,height=3.01in,width=3.01in}}
\vspace{10pt}
\caption{
News from three different runs: $\bigtriangleup$ marks the results for
 $A=1$, $q=0$ (multiplied by 25 to compensate linear scaling),
the solid line denotes $A=5$, $q=0$ and
$\diamond$ the run for $A=5$, $q=-r^2/a$.}
\label{fig:News}
\end{figure}

Fig. \ref{fig:News} shows the news function from three different runs:
for $A=1,5$ and $q=0$, and for $A=5$, $q=-r^2/a$. The news from the
runs for $A=1$ have been multiplied by a factor of 25, which would exactly
compensate scaling with the amplitude in the linear regime.
We see that the three curves line up very well initially. The line
for $A=5$ and $q=0$ deviates significantly to larger values of the news
when the runs starts to get inaccurate, but at this time most of the
physical radiation has already left the system.
The curves from  $A=1$/$q=0$ and $A=5$/$q=-r^2/a$ line up perfectly until
the value of the news drops below $10^{-6}$, where the curves level off
at different values, due to numerical inaccuracy.

\begin{figure}[htbp] % [b!] fig 1
\centerline{\epsfig{file=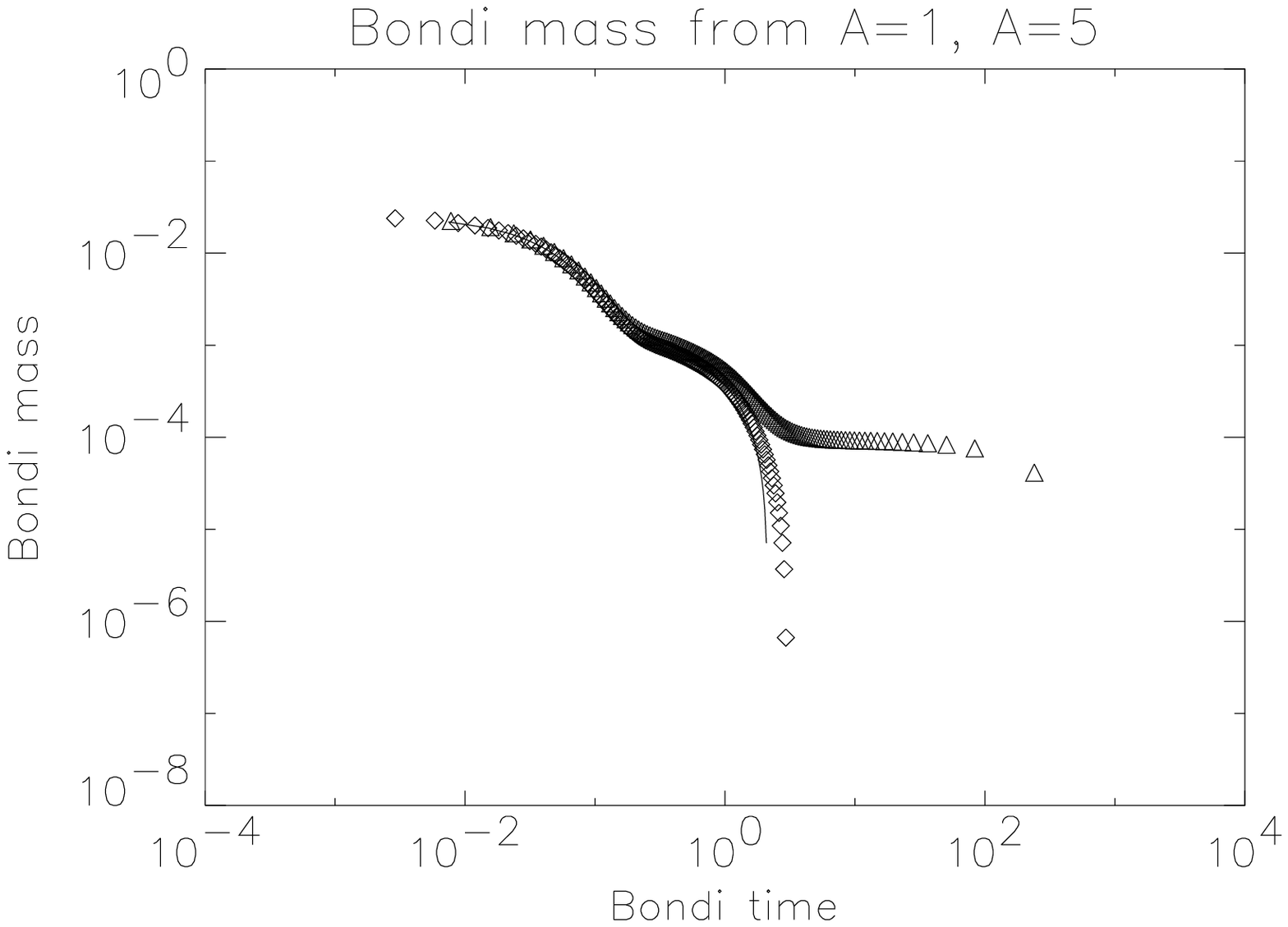,height=3.01in,width=3.01in}}
\vspace{10pt}
\caption{
Bondi mass from three different runs: $\bigtriangleup$ marks the results for
 $A=1$, $q=0$ (multiplied by 25 to compensate linear scaling), the solid line denotes $A=5$, $q=0$ and
$\diamond$ the run for $A=5$, $q=-r^2/a$.}
\label{fig:BondiMass}
\end{figure}

Fig. \ref{fig:BondiMass} shows the Bondi mass for this situation, again the
$A=1$ curve is scaled by a factor of 25:
Again we see the quick decay of a sharp pulse of
radiation. There is no particular structure except falloff
at late times, the deviation of the curves at late times seems to be caused
by numerical inaccuracy, in particular in the computation of the Bondi mass.

\section{Computational aspects}\label{sec:computational}
%%%%%%%%%%%%%%%%%%%%%%%%%%%%%%%

In this section I will give a brief description of some computational
aspects, such as the computational
resources needed to carry out simulations in 3 spatial dimensions.
Computations of this scale rely on parallel processing, that means execution
of our algorithms is
spread over different CPU's. From a simplistic point of view there are two ways
to program for parallel execution: we only take care of parallelizing the
algorithm -- but we require that all CPU's can access the same memory -- or we
both parallelize the algorithm and the data structures, and separate
the total  data into smaller chunks that fit into the local memory of
each processor. The first alternative requires so called shared memory
machines, where the operating system and hardware take care of making data
accessible to the CPU's consistently, taking care of several layers of main
and cache memory
(which gets increasingly difficult and expensive as the size of the
machine is increased). 
The present code has been implemented using a shared memory programming model.
The advantage is that this can generally be somewhat easier to program,
and avoids overheads in memory. The disadvantage is the high cost of such
systems, which makes them difficult to afford and thus nonstandard for
most large academic parallel applications. The second alternative, usually
referred to as distributed memory, requires more work to be done
by the programmer,
but more flexibly adapts to different kinds of machines such as clusters
of cheap workstations commonly available in academic environments.
While this approach usually implies a larger overhead in total memory 
requirements, speed and programming complexity, it is currently the only
approach capable of scaling from small to very large simulations.
For a general introduction to the issues of high performance computing,
Ref. \cite{OreillyHPC} provides a good starting point.

So, how much memory do we need?
Let us assume a run with $150^3$ grid points
(the size of the largest simulations
carried out with the present code so far). The current implementation
of the fourth order Runge-Kutta
algorithm uses 4 time levels and a minimum of 62 grid functions
(57 variables and 5 gauge source functions). In double precision this amounts
to $\approx 6.7$ GByte. Temporary variables, information on geodesics
and various overheads result in a typical increase of memory
requirements by roughly $20 \%$.
For 150 time steps (approximately what it takes to reach $i^+$ for weak data)
the total amount of processed data then corresponds to roughly 1 Terabyte!

If we half the grid spacing, the allocated memory increases by a factor of
$8$ (neglecting overheads), the total amount of processed 
data by a factor of $16$, and the total required CPU time also
by a factor of $16$, while the error {\em reduces} by a factor of
$16$  -- {\bf if} we are already in the convergent regime!
Given that the  biggest academic shared memory machines in Germany
have 16 GByte of memory available (the AEI's Origin 2000 and the Hitachi
SR8000  at LRZ in Munich) this shows that the margin for increasing resolution
is currently quite small. Such an increase in resolution will however be
necessary to resolve physically interesting situations with more structure,
such as a black hole, or two merging black holes.
A move toward distributed memory processing will therefore be
likely in the long run.

The current software-standard for distributed (scalable) computing is
MPI (message passing interface) \cite{MPI}.
Unfortunately, writing large scale sophisticated 
codes in MPI is very time consuming.
However, several software packages are available which introduce 
a software layer between the application programmer and MPI, and thus
significantly reduce the effort to write parallel applications.
Two prime examples are the Cactus Computational Toolkit \cite{Cactus}
and PETSc \cite{PETSc}. 
While  PETSc is a general purpose tool developed at Argonne National Laboratory
as parallel framework for numerical computations, Cactus has been
developed at the Albert Einstein Institute with numerical relativity in mind.
While PETSc offers more support for numerical algorithms, in particular
for parallel elliptic solvers, Cactus already contains some general numerical
relativity functionality like apparent horizon finders -- but no support for
generic numerical algorithms.
Apart from its numerical relativity flavor, the Cactus computational toolkit
also has the advantage of broad support for parallel I/O and large scale
3D visualization. The ability to successfully mine
tens or hundreds of gigabytes of data for relevant features is paramount
to successful simulations in 3D.
 
Among the essential problems in writing and maintaining large scientific
codes are the software engineering aspects and the control of complexity.
In other words, codes should be reasonably documented and maintainable.
For large scientific codes written and maintained by part-time-programmer
scientists this poses a significant challenge. Writing
a clear, modular code that can be understood, maintained and extended 
to suit new scientific needs requires a good deal of design and planning
ahead. For an introduction
to software engineering issues see e.g. \cite{MMM}, \cite{code_complete}.
Another important issue for scientific codes is flexibility. Being able to
do good science often depends on the ability to easily change algorithms,
equations, discretization schemes etc. -- without having to restructure the
code, without high risk of introducing new bugs.
In the present case examples for the need of trying different
things would be experiments with different evolution equations (e.g. metric
versus frame formalism), different boundary treatments or different elliptic
solvers.

\section{Discussion}\label{sec:discussion}
%%%%%%%%%%%%%%%%%%%%%

The 3D numerical simulations performed so far show that the
evolutions are {\em numerically} stable and quite robust. 
However, one of the main problems in numerical relativity is the stability
of the constraint propagation: while the constraints {\em do} propagate when
they are satisfied identically initially, this assumption does not
hold for numerical simulations. On the contrary -- it seems to be quite typical
observation that the constraints diverge exponentially, if
the evolution does not start at the constraint surface.
Preliminary results exhibit this behavior
of resolution-independent exponential growth associated with a violation
of the constraints also for the conformal approach. One of the major
goals for the future thus has to be the improvement of the understanding
of the constraint propagation equations, and an according modification
of the evolution equations (see Ref. \cite{control_constraints}
for previous work in this direction). This is essentially an analytical
problem, but will certainly require the numerical testing of ideas.

Another area where new developments are necessary on the analytical side --
along with numerical testing -- is the problem of finding gauges that
prevent pathologies like unnecessarily strong gradients.
Ideally one would want to keep the symmetric hyperbolic character of the
evolution system while allowing for a maximum of flexibility in writing down
evolution equations for the gauge source functions.   

Well-posedness of the evolution equations is important -- but by far not
sufficient for numerical purposes. While well-posedness unfortunately
still has not yet been shown rigorously for many formulations used in
numerical relativity, another important task seems to be to improve
the understanding of the non-principal part of the equations,
including their nonlinearities, in order to be able to
construct numerically well-conditioned algorithms.

The third area where significant progress seems necessary on the analytical
side is the construction of initial data. Problems with the
current algorithm which
necessitates divisions by zero and an ad-hoc extension beyond $\scri$
have not yet been resolved. A possible road toward resolving these problems
has been outlined by Butscher in this volume.

An important role in improving the analytical understanding and in setting
up numerical experiments will be played by the utilization of
simplifications. Particularly important are spacetime symmetries
and perturbative studies.
A particularly interesting case to be studied actually is Minkowski space.
Besides being an important case for code testing, it is used in
current investigations to learn more about gauges and the stability of
constraint propagation.

More complicated are general spherically symmetric spacetimes. In the vacuum
case, this only leaves the Kruskal spacetime aside from Minkowski space --
but understanding the gauge problem for Kruskal spacetime is an important
milestone toward long-time black hole simulations.
Moreover, spherical symmetry provides a natural testing ground
for all kinds of new ideas, e.g. of how to treat the appearance of
singularities, of how to treat the unphysical region,
numerical methods, etc.

An alternative route to simplification, which has been very successful
in numerical relativity, is perturbative analysis, e.g. with Minkowski
or Schwarzschild backgrounds. In the context of compactification
this has been carried out numerically
with characteristic codes in \cite{pertI,pertII}, some of the problems
that showed up there are likely to be relevant also for the conformal approach.

What can we expect from the conformal approach in terms of
physics results? Where can we expect contributions to our understanding
of general relativity?
One of the most important features of the conformal approach is that
it excels at radiation extraction without ambiguities, and -- at least
in principle -- enables numerical codes to study the global
structure of spacetimes describing isolated systems. 
As has been demonstrated in this paper, in some weak field regime the
code works well with relatively simple choices of gauge, and could be used
to investigate some of the above problems.
It could also provide a very clean way
to study nonlinear deviations from linear predictions.
For strong fields, in particular
one or two black holes, the problem is much more difficult.
An even more difficult problem is the investigation of the structure of
singularities. In the spherically symmetric case this has been
achieved by H\"ubner \cite{Hu96mf},
but it is not clear whether these methods can be carried over to
the generic case without symmetries, where the structure of the singularity
has to be expected to be much more complicated.

%%\section{Conclusions}\label{sec:conclusions}
%%%%%%%%%%%%%%%%%%%%%%%

What is the roadmap for the future?
As far as 3D simulations are concerned, I believe that
one should try to go from relatively well controlled weak data to stronger
data and try to identify and solve problems as they come up.
%Main aims are to get more experience with gauges, and to demonstrate
%that we can evolve in the strong field regime, i.e. observe
%the appearance of an apparent horizon, see a strong news function come out of
%the system, etc.
In parallel, it will be important to study simplified situations,
like spacetimes with symmetries or linear perturbations with a mixture
of analytical and numerical techniques. 
Both lines of research are hoped to improve our understanding of
issues associated with choosing the gauge source functions,
and controlling the growth of constraints.
Future 3D codes, aimed at producing novel physical results
will also require
a significant effort devoted to ``computational engineering'', since
flexible and solidly written codes are an absolute
necessity for good computational science!
These well known problems plaguing 3D numerical relativity will have to be
addressed and solved in the conformal approach in order to harvest its
benefits.
Developing the conformal approach to numerical
relativity into a mature tool poses an important challenge for
mathematical relativity: not only is the problem hard and requires long-term
investments, it also requires to merge sophisticated mathematical analysis with
computational engineering.
The aim is to produce a solid handle on exciting new physics
-- and some of the physics will even be accessible to experiments.

\section*{Acknowledgments}
The author thanks H. Friedrich, B. Schmidt, M. Weaver and J. Frauendiener
for helpful discussions and explanations of their work,
C. Lechner and J. Thornburg for a careful reading
of the manuscript, and P. H\"ubner for giving me access to his codes and
results, and for support in the early stages of my work on this subject.

%INDEX%%%%%%%%%%%%%%%%%%%%%%%%%%%%%%%%%%%%%%%%%%%%%%%%%%%%%%%%%%%%%%%
% Please check with the editor of your book whether he plans to
% include a "mutual" subject index - if so, please code your entries
% in the standard syntax. For your own purposes you may print your
% "personal" index by using the following commands:
%
%\clearpage
%\addcontentsline{toc}{section}{Index}
%\flushbottom
%\printindex
%%%%%%%%%%%%%%%%%%%%%%%%%%%%%%%%%%%%%%%%%%%%%%%%%%%%%%%%%%%%%%%%%%%%%


\begin{thebibliography}{8.}
\addcontentsline{toc}{section}{References}
\bibitem{hahnlindquist64}
S. Hahn and R. Lindquist, Annals of Physics 29, (1964) 304.

\bibitem{smarr78}
L. Smarr, in {\it Sources of gravitational radiation\/}, ed. L. Smarr,
Seattle, Cambridge University Press, 1978.

\bibitem{eppley77}
K. Eppley, Phys. Rev. D 16 (1977) 1609.

\bibitem{lazarus}
J. Baker, B. Br\"ugmann, M. Campanelli and C.~O. Lousto, Class. Quant.
  Grav. 17 (2000) L149.

\bibitem{Wald}
R. Wald, {\it General Relativity\/}, University of Chicago Press, 
Chicago, 1984.

\bibitem{joerg_livrev}
J. Frauendiener, Living Rev. Rel. 4 (2000) 1.

\bibitem{HuAustin}
S. Husa, in ``Proceedings of the 20th Texas Symposium on
Relativistic Astrophysics'', ed. by J. C. Wheeler and H. Martel,
American Institute of Physics, 2001.

\bibitem{PeterI}
P.~H\"ubner,
%,How to Avoid Artificial Boundaries in the Numerical Calculation of Black
% Hole Spacetimes
Class. Quant. Grav. 16 (1999) 2145--2164.

\bibitem{PeterII}
P.~H\"ubner,
% A Scheme to Numerically Evolve Data for the Conformal Einstein Equation
Class. Quant. Grav. 16 (1999) 2823--2843.

\bibitem{PeterIII}
P.~H\"ubner,
% Numerical Calculation of Conformally Smooth Hyperboloidal Data
Class. Quant. Grav. 18 (2001) 1421--1440.

\bibitem{PeterIV}
P.~H\"ubner,
%``From now to timelike infinity on a finite grid,''
Class. Quant. Grav.  18 (2001) 1871--1884.
%%CITATION = GR-QC 0010069;%%

\bibitem{ScriFixing}
J. Frauendiener, 
% Numerical treatment of the hyperboloidal initial value problem for
% the vacuum Einstein equations. II. The evolution equations.
Phys. Rev. D 58 (1998) 064003.

\bibitem{Fr86ot}
H.~Friedrich, Commun. Math. Phys. 107 (1986) 587--609.

\bibitem{Pe63ap}
R.~Penrose, Phys. Rev. Lett. 10 (1963) 66--68.

\bibitem{cosmic-censorship}
R.~Penrose,
Rivista del Nuovo Cimento 1 (1969) 252;
%
R.~Penrose,
%"The Question of Cosmic Censorship",
in {\em Black Holes and Relativistic Stars}, ed. by R. Wald,
(Chicago University Press, Chicago, 1988).

\bibitem{AnCA92ot}
L.~Andersson, P.~T. Chrusc\'{\i}el, H.~Friedrich,
Comm. Math. Phys. 149 (1992) 587.

\bibitem{Fornberg}
B. Fornberg,
{\em A Practical Guide to Pseudospectral Methods}, 
Cambridge University Press, Cambridge, 1996.

\bibitem{Canuto}
C. Canuto, M. Hussaini, A. Quarteroni and T. Zang, {\em Spectral Methods in
Fluid Dynamics\/}, Springer Verlag, New York, 1988.

\bibitem{Boyd}
J. Boyd, {\em Chebyshev \& Fourier Spectral Methods\/},
Springer Verlag, Berlin, 1989.

\bibitem{FFTW}
M.~Frigo and S.~G. Johnson,
``Fftw: An adaptive software architecture for the fft'' in
{\em 1998 ICASSP conference proceedings, Vol. 3}, ICASSP, 1998, p. 1381.

\bibitem{NewtonIteration}
J.  Stoer and R.  Burlisch,
{\em  Introduction to Numerical Analysis. 2nd edition},
Springer, Berlin, 1996.

\bibitem{AMG}
J.~W. Ruge and K.~St\"uben, ``Algebraic multigrid'', in
 {\em Multigrid Methods}, edited by
 S.~F. McCormick, SIAM, Philadelphia, pp. 73--130, 1987;
K.~St{\"u}ben, {\em GMD report}, 1999.

\bibitem{FriedrichOverview}
H.~Friedrich,  
%``Einstein's Equation and Geometric Asymptotics'',
\newblock in {\em Proceedings of the GR-15 conference}, edited by
N. Dadhich and J. Narlikar, IUCAA, 1998.

\bibitem{QuV97NA}
A.~Quarteroni and A.~Valli,
 {\em Numerical Approximation of Partial Differential Equations},
 Springer Series in Computational Mathematics 23, Springer, Berlin, 1997.

\bibitem{GuKA95TD}
B.~Gustafsson, H.-O. Kreiss, and J.~Oliger,
``{\em Time Dependent Problems and Difference Methods}'',
Pure and Applied Mathematics. Wiley, New York, 1995.

\bibitem{bernd_kruskal}
B. Schmidt, in ``Proceedings of the 20th Texas Symposium on
Relativistic Astrophysics'', ed. by J. C. Wheeler and H. Martel,
American Institute of Physics, 2001.

\bibitem{BeOM}
R. Beig and N. \'O Murchadha Phys. Rev. Lett. 66 (1991) 2421.

\bibitem{Be94Ini}
R. Beig and S.~Husa,
% Initial Data for General Relativity with Toroidal Conformal Symmetry
Phys. Rev. D 50 (1994) 7116-7118.

\bibitem{HuDiss}
S.~Husa,
``{\em Asymptotically Flat Initial Data for Gravitational Wave Spacetimes, Conformal Compactification and Conformal Symmetry}'',
PhD thesis, University of Vienna 1998.

\bibitem{bernd_punctures}
S. Brandt and B. Br\"ugmann,
% A simple construction of initial data for multiple black holes
Phys. Rev. Lett. 78 (1997) 3606-3609.

\bibitem{Dain}
S.~Dain, Phys. Rev. Lett. 87 (2001) 121102.

\bibitem{Miguel}
M. Alcubierre and B. Br\"ugmann,
% Simple excision of a black hole in 3+1 numerical relativity
Phys. Rev. D 63 (2001) 104006.

\bibitem{control_constraints}
 O. Brodbeck, S. Frittelli, P. H\"ubner and O. Reula,
   J. Math. Phys. 40 (1999) 909--923;
%
 M. Alcubierre, G. Allen, B. Bruegmann and E. Seidel,
   Wai-Mo Suen, Phys. Rev. D62 (2000) 124011;
%
 G. Yoneda and H. Shinkai, Class. Quant. Grav. 18 (2001) 441--462;
%
 F. Siebel and P. H\"ubner, Phys. Rev. D64 (2001) 024021.

\bibitem{Hu96mf}
P.~H\"ubner,
\newblock {\em Phys. Rev. D} {\bf 53}, 701--721 (1996).

\bibitem{pertI}
M. Campanelli, R. Gomez, S. Husa, J. Winicour, Y. Zlochower,
% The close limit from a null point of view: the advanced solution
Phys. Rev. D 63 (2001) 124013.

\bibitem{pertII} 
S. Husa, Y. Zlochower, R. Gomez, J. Winicour,
% Retarded radiation from colliding black holes in the close limit
 {\em Los Alamos Preprint Archive}, gr-qc/0108075, to appear
in Phys. Rev. D (2002).

\bibitem{OreillyHPC}
K.  Dowd, C. R. Severance,
{\em  High Performance Computing}, 2nd Ed., O'Reilly, Cambridge, 1998.

\bibitem{Cactus}
http://www.cactuscode.org.

\bibitem{PETSc}
S. Balay, W. D. Gropp, L. Curfman McInnes and B. F. Smith,
 "{PETSc} Users Manual", Technical Report ANL-95/11, 2001; 
%
S. Balay, W. D. Gropp, L. Curfman McInnes and B. F. Smith,
"Efficient Management of Parallelism in Object Oriented Numerical Software Libraries'', in Modern Software Tools in Scientific Computing, ed. by
E. Arge and A. M. Bruaset and H. P. Langtangen, Birkhauser Press, 1997;
http://www.mcs.anl.gov/petsc.

\bibitem{MPI}
W.  Gropp, E.  Lusk and A. Skjellum, 
{\em  Using MPI}, 2nd Ed., MIT Press, Cambridge, Massachusetts,
1999.

\bibitem{MMM}
F. P.  Brooks Jr.,
{\em The Mythical Man-Month}, Addison-Wesley, Reading, 1995.

\bibitem{code_complete}
S.  McConnell, {\em  Code Complete -
 A Practical Handbook of Software Construction}, Microsoft Press, 
Redmond, 1993.


\end{thebibliography}
\end{document}